\documentclass[lettersize, conference]{IEEEtran}
\usepackage{cite}
\usepackage{amsmath,amssymb,amsfonts}
\usepackage{algorithmic}
\usepackage{graphicx}
\usepackage{textcomp}
\usepackage{xcolor}
\usepackage{dsfont}
\usepackage{subfig}
\usepackage{float}
\def\BibTeX{{\rm B\kern-.05em{\sc i\kern-.025em b}\kern-.08em
    T\kern-.1667em\lower.7ex\hbox{E}\kern-.125emX}}

\usepackage[colorlinks=true, urlcolor=blue, citecolor=blue]{hyperref}

\newtheorem{theorem}{Theorem}[section]

\newtheorem{observation}[theorem]{Observation}

\usepackage[T1]{fontenc}
\usepackage[utf8]{inputenc}
\usepackage{chronosys}

\usepackage[ruled,vlined]{algorithm2e}
\usepackage{listings}
\definecolor{codegreen}{rgb}{0,0.7,0.1}
\definecolor{codegray}{rgb}{0.5,0.5,0.5}
\definecolor{codepurple}{rgb}{0.58,0,0.82}
\definecolor{backcolour}{rgb}{0.9,0.9,0.9}
 
\lstdefinestyle{mystyle}{
    backgroundcolor=\color{backcolour},   
    commentstyle=\color{codegreen},
    keywordstyle=\color{magenta},
    numberstyle=\tiny\color{codegray},
    stringstyle=\color{codepurple},
    basicstyle=\footnotesize,
    breakatwhitespace=false,         
    breaklines=true,                 
    captionpos=b,                    
    keepspaces=true,                 
    numbers=left,                    
    numbersep=5pt,                  
    showspaces=false,                
    showstringspaces=false,
    showtabs=false,                  
    tabsize=2
}
\newcommand{\LDPC}{LDPC }

\title{A mapping of the Min-Sum decoder to reduction operations, and its implementation using CUDA kernels}

\makeatletter
\newcommand{\linebreakand}{%
  \end{@IEEEauthorhalign}
  \hfill\mbox{}\par
  \mbox{}\hfill\begin{@IEEEauthorhalign}
}
\makeatother

\author{
  \IEEEauthorblockN{Omer S. Sella}
  \IEEEauthorblockA{\textit{dept. of computing} \\
    \textit{Imperial College London}\\
    o.sella@imperial.ac.uk}
  \and
  \IEEEauthorblockN{Thomas Heinis}
  \IEEEauthorblockA{\textit{dept. of computing} \\
    \textit{Imperial College London}\\
    t.heinis@imperial.ac.uk}
}

\begin{document}
\lstset{language=Python}
\maketitle
\begin{abstract}
Decoders for Low Density Parity Check (LDPC) codes are usually tailored to an application and optimized once the specific content and structure of the parity matrix are known. In this work we consider the parity matrix as an argument of the Min-Sum decoder, and provide a GPU implementation that is independent of the content of the parity matrix, and relies only on its dimensions.
\end{abstract}
\section{Introduction}
\LDPC codes made their way into several standards and technologies, spanning telecommunications and storage \cite{anderson2023project}. 
Prior to adoption, it is sometimes the case that several codes are considered, stemming from different constructions or parameter tuning of a single construction \cite{Thorpe2003}. In this early, short-lived stage, decoder optimizations that are a function of the specific parity matrix are not readily available. In this stage, a high throughput decoder helps eliminate or choose from a selection of codes, often having a parity matrix of similar dimensions.
This work is concerned with the optimization of the Min-Sum decoder when only the dimensions of the parity matrix $H$ are known.
\section{Preliminaries\label{sec:preliminaries}}
The material presented here is entirely covered or can be deduced from prior literature. Nevertheless, we believe that the practical significance of it makes it worthwhile to present here.
\subsection{A collection of linear equations over a binary field}
\begin{figure}[h]
	\centering
	\includegraphics[width=0.2\linewidth]{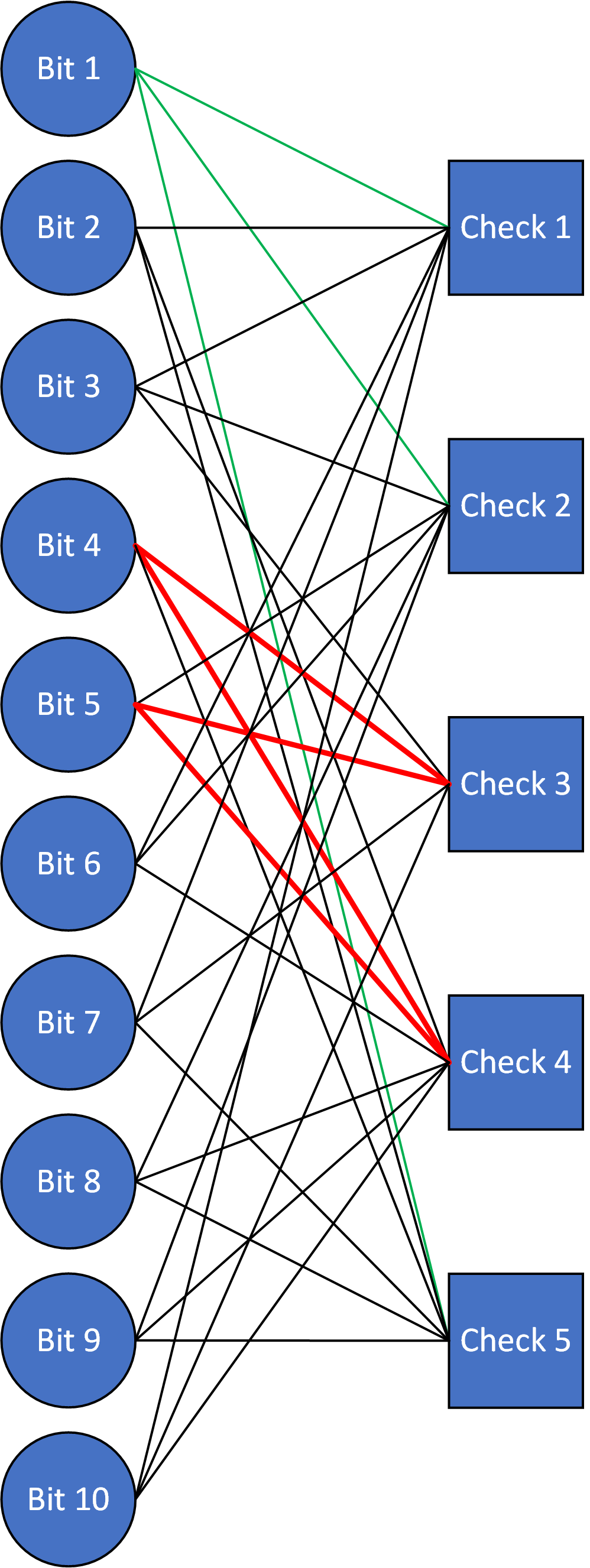}
	\caption{Tanner graph example. \label{fig:TannerSmall}}
\end{figure}

The reader may be familiar with the term Single Parity Checksum (SPC). 
Given a vector $v$ from the $k$ dimensional space $\{0,1\}^k$ over $F_2$, 
an SPC is defined by a choice $I \subset \{1,..,k\}$ of indices from the set of integers $\{1,..,k\}$.
An SPC is said to be satisfied if 
\begin{equation}
   \sum_{i \in I} v(i) = 0 \quad (mod\,2)  
\end{equation}
which is identical to the bit-wise XOR of the values of the specific coordinates being $0$, i.e.: 
\begin{equation}
  \oplus_{i\in I} v(i) = 0
\end{equation}
\begin{figure}[h]
	\centering
    \includegraphics[width=0.2\textwidth]{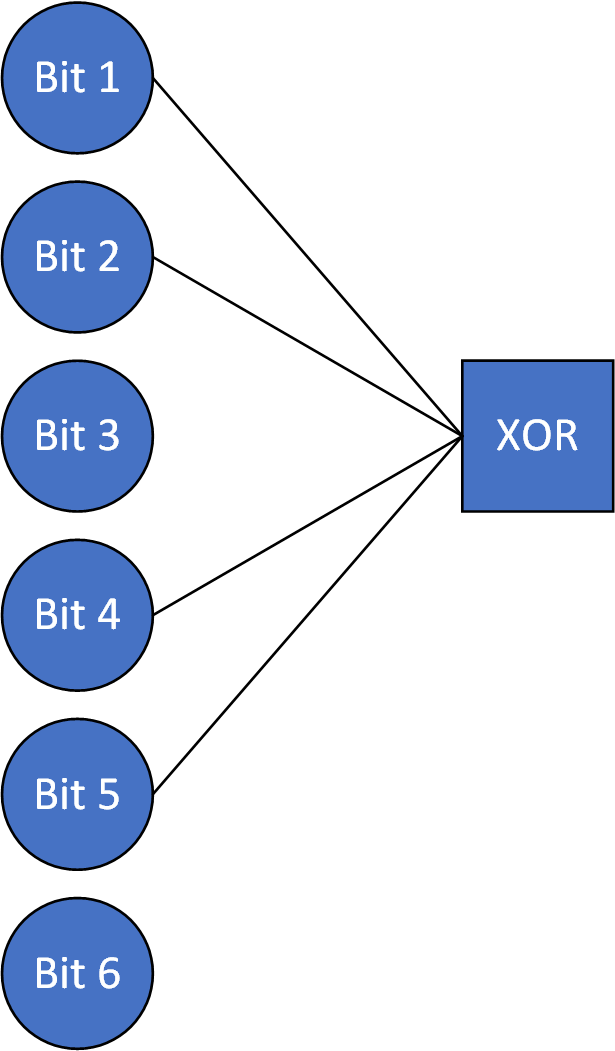}
	\caption{A single parity checksum. Bits 1 through 6 may take any value in ${0,1}$. 
We say that the single parity checksum is satisfied if 
$0 = Bit_1 \oplus Bit_2 \oplus Bit_4 \oplus Bit_5$ 
corresponding to the index set $I=\{1,2,4,5\}$. 
The corresponding vector $\mathds{1}_I = (1,1,0,1,1,0)$ \label{fig:SPC}}
\end{figure}
The SPC pictured in Figure~\ref{fig:SPC} corresponds to the case where $k=6$ and $I=\{1,2,4,5\}$. Bipartite graphs can be viewed as a collection of SPCs. The bipartite graph in Figure~\ref{fig:TannerSmall} is also known as a Tanner graph \cite{Tanner1981} or a Forney Factor Graph (FFG) \cite{Loeliger2004}. The nodes in the bipartite graph defined by H (i.e., the Tanner graph) are labelled as two enumerated sets $B, C$. 
The nodes in the set $C$ represent the check nodes $c_1, c_2, ... $ shown on the right side of Figure~\ref{fig:TannerSmall}
that impose relations on the circular bit nodes from the set $B$.
We may observe that this graph could be viewed as a collection of the following SPCs:
\begin{itemize}
    \item Check 1 is defined by $$I_1 = \{1,2,3,6,7,10\}$$ and yields $\mathds{1}_{I_1} = (1,1,1,0,0,1,1,0,0,1)$
    \item Check 2 is defined by $$I_2 = \{1,3,5,6,8,9\}$$ and yields $\mathds{1}_{I_2} = (1,0,1,0,1,1,0,1,1,0)$
    \item Check 3 is defined by $$I_3 = \{3,4,5,7,9,10\}$$ and yields $\mathds{1}_{I_3} = (1,1,1,0,0,1,1,0,0,1)$
    \item Check 4 is defined by $$I_4 = \{2,4,5,6,8,10\}$$ and yields $\mathds{1}_{I_4} = (0,1,0,1,1,1,0,1,0,1)$
    \item Check 5 is defined by $$I_5 = \{1,2,4,7,8,9\}$$ and yields $\mathds{1}_{I_5} = (1,1,0,1,0,0,1,1,1,0)$
\end{itemize}
A binary vector $v \in \{0,1\}^{10}$ satisfies all these SPCs concurrently if and only if it is in the kernel
of all the vectors above, i.e., $v \in ker(H)$ where the rows of $H$ are exactly the above vectors.

\subsection{The Min-Sum algorithm}
\begin{figure}[h]
  \centering
\subfloat[A binary matrix. The check nodes are enumerated according to the row numbers and are presented here in the leftmost column. The bit nodes are enumerated according to the column numbers and are presented here on the top row.
\label{subfig:parityMatrix}]{\includegraphics[width=0.3\textwidth]{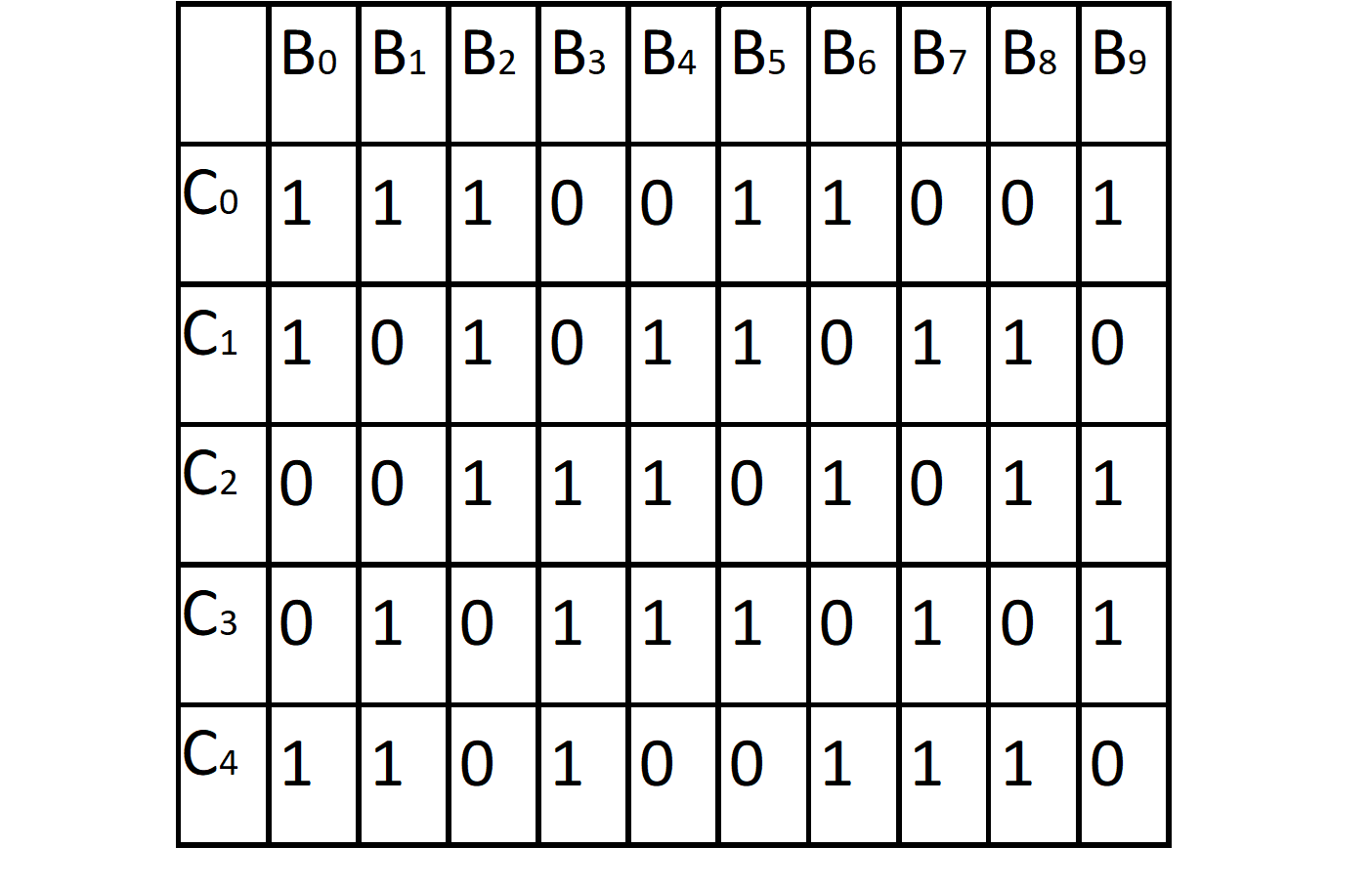}}
\hspace{0.5cm}
\subfloat[The indices corresponding to of In this example we show how to find the set of bit nodes $N_4$. We start at the row corresponding to \textit{check node} number $4$, i.e.: $c_4$ and find the non-zero elements in that row.]{%
      \includegraphics[width=0.3\textwidth]{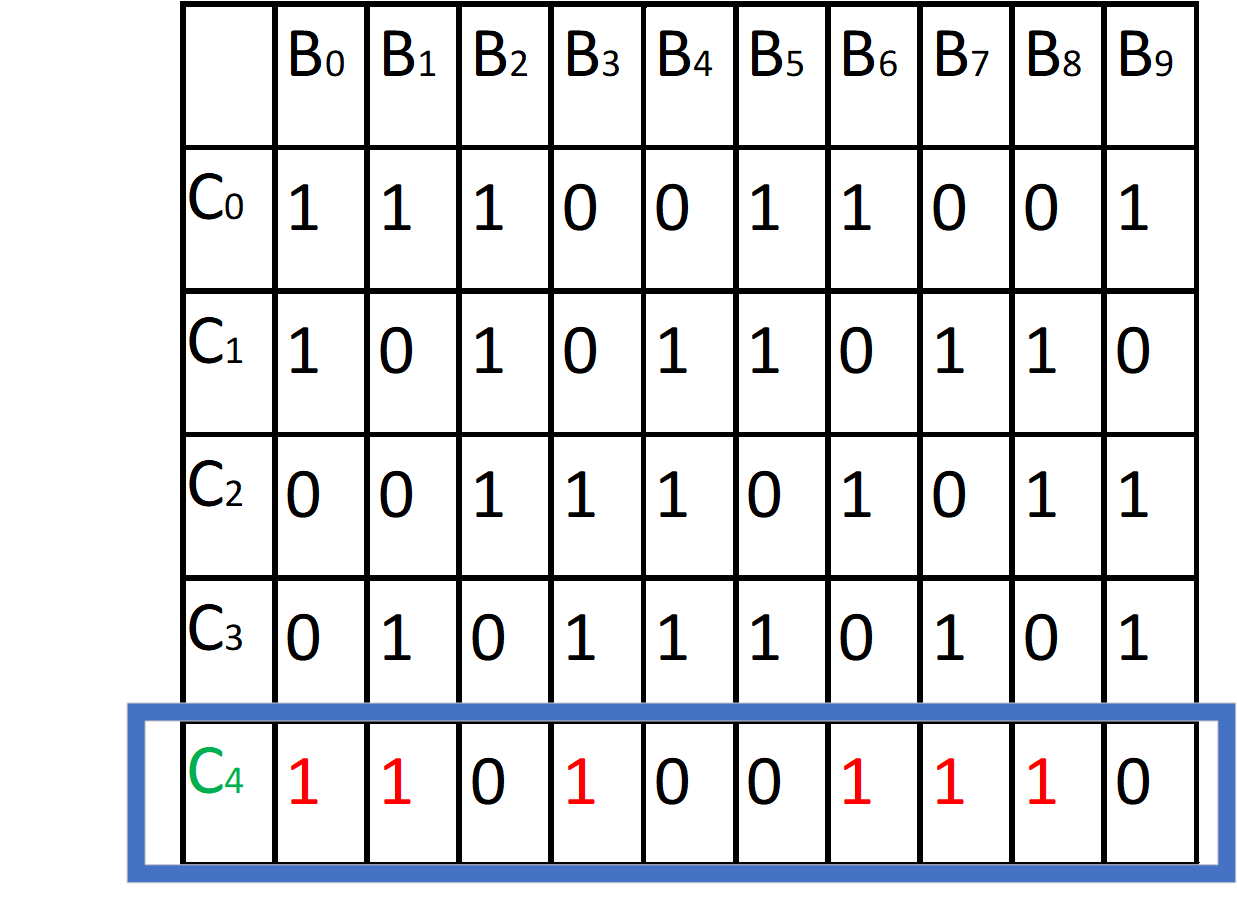}}
\hspace{0.5cm}
\subfloat[The location of each non-zero element found in the row indicates a \textit{bit node} that belongs to the set $N_4$.]{%
      \includegraphics[width=0.3\textwidth]{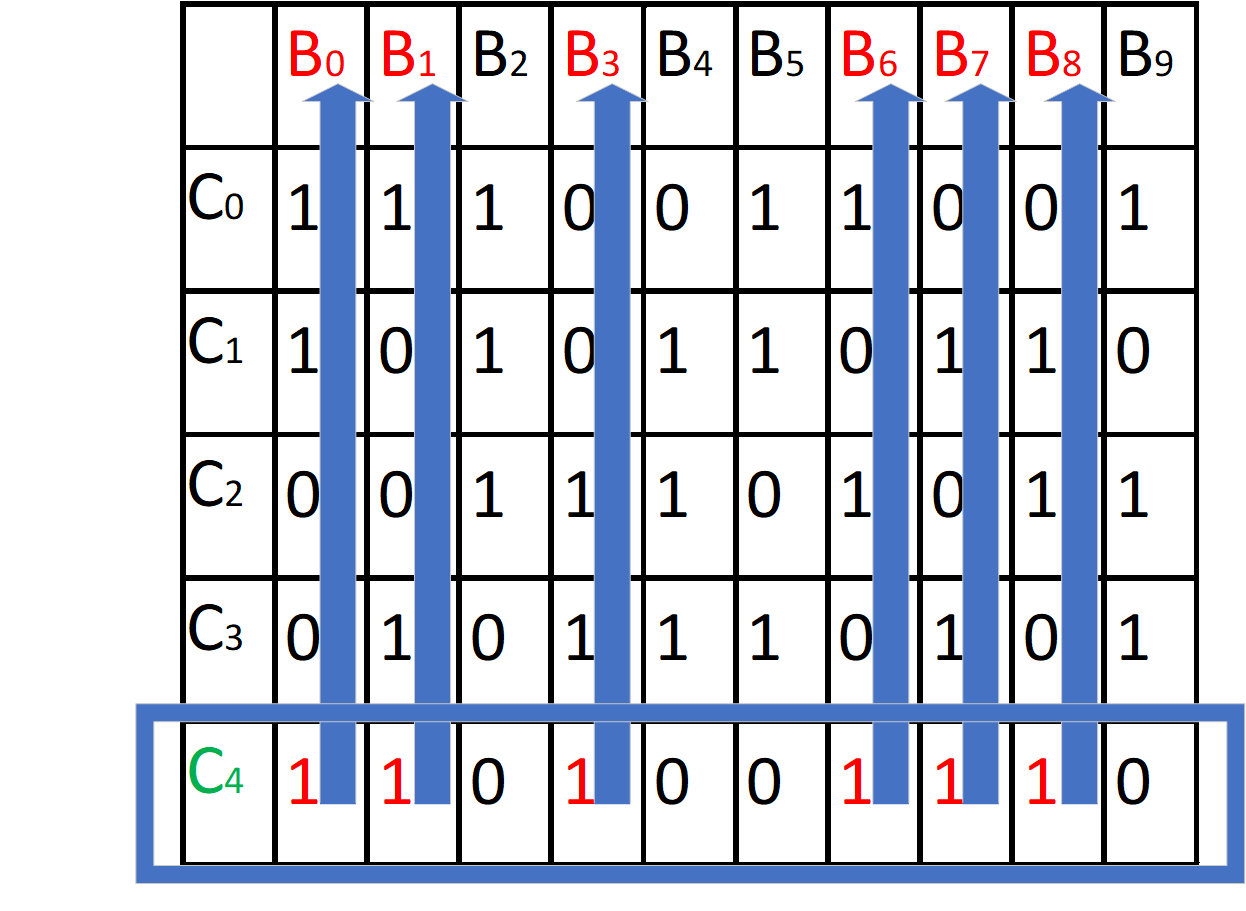}}
\caption{The sets $N_i$ are defined for every check node and so correspond to the locations of non-zero elements in the rows of the parity matrix.}
\label{fig:understandingNi} 
\end{figure}

\begin{figure}[ht]
  \centering
\subfloat[We use the same binary matrix as in Figure~\ref{fig:understandingNi}]
{\includegraphics[width=0.3\textwidth]{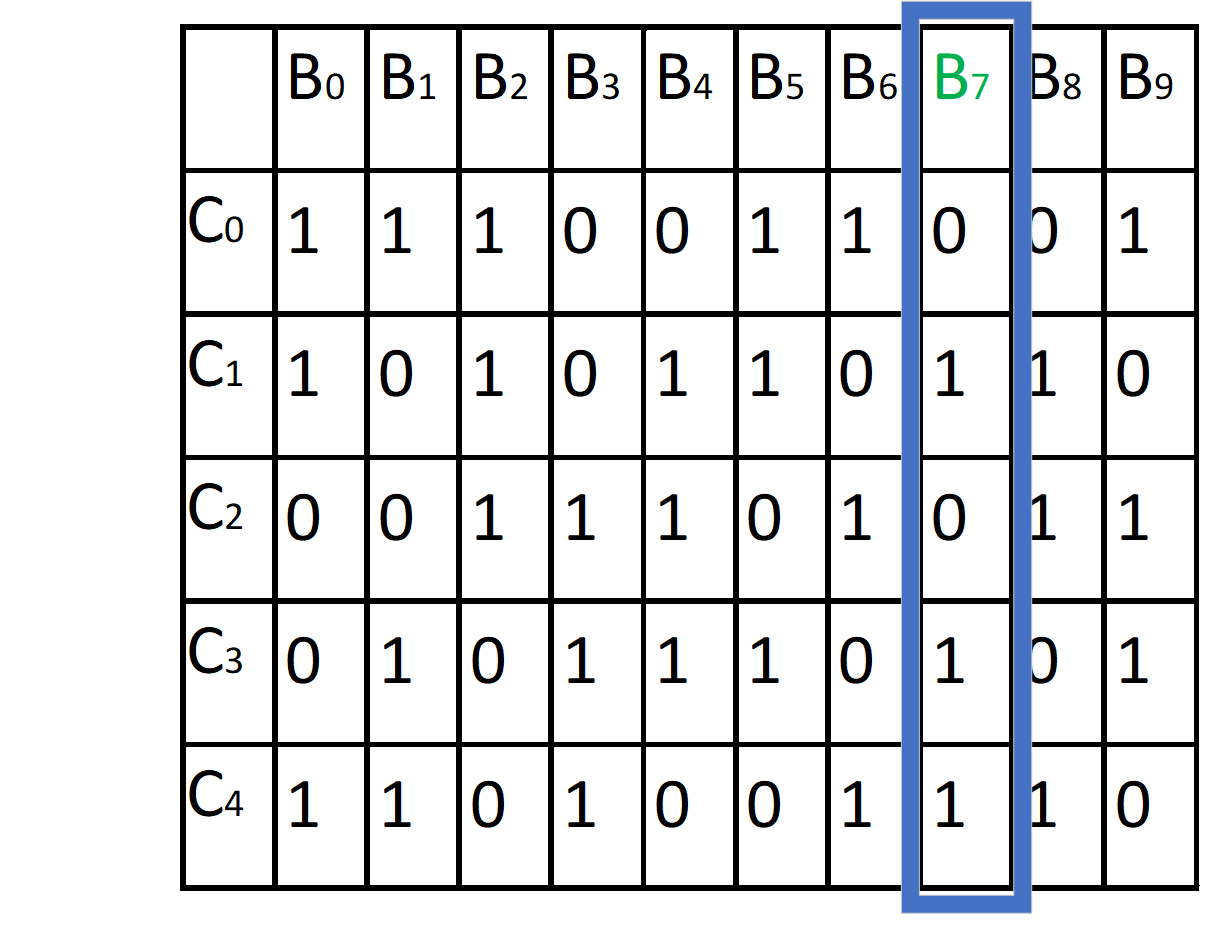}}
\hspace{0.5cm}
\subfloat[In this example we show how to find the set of bit nodes $M_7$. We start at the row corresponding to \textit{bit node} number $7$, i.e.: $B_7$ and find the non-zero elements in that row.]{%
      \includegraphics[width=0.3\textwidth]{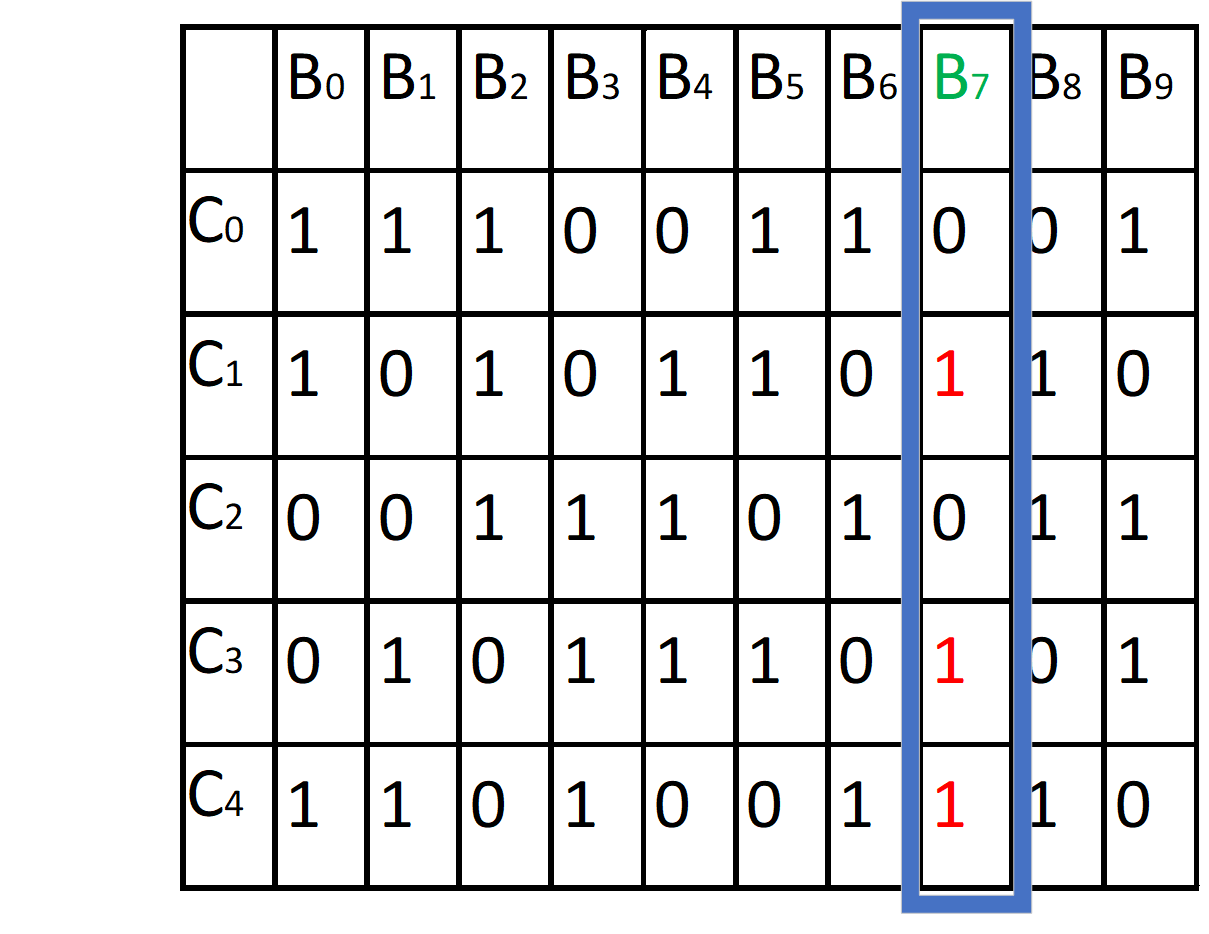}}
\hspace{0.5cm}
\subfloat[The location of each non-zero element found in the row indicates a \textit{check node} that belongs to the set $M_7$.]{%
      \includegraphics[width=0.3\textwidth]{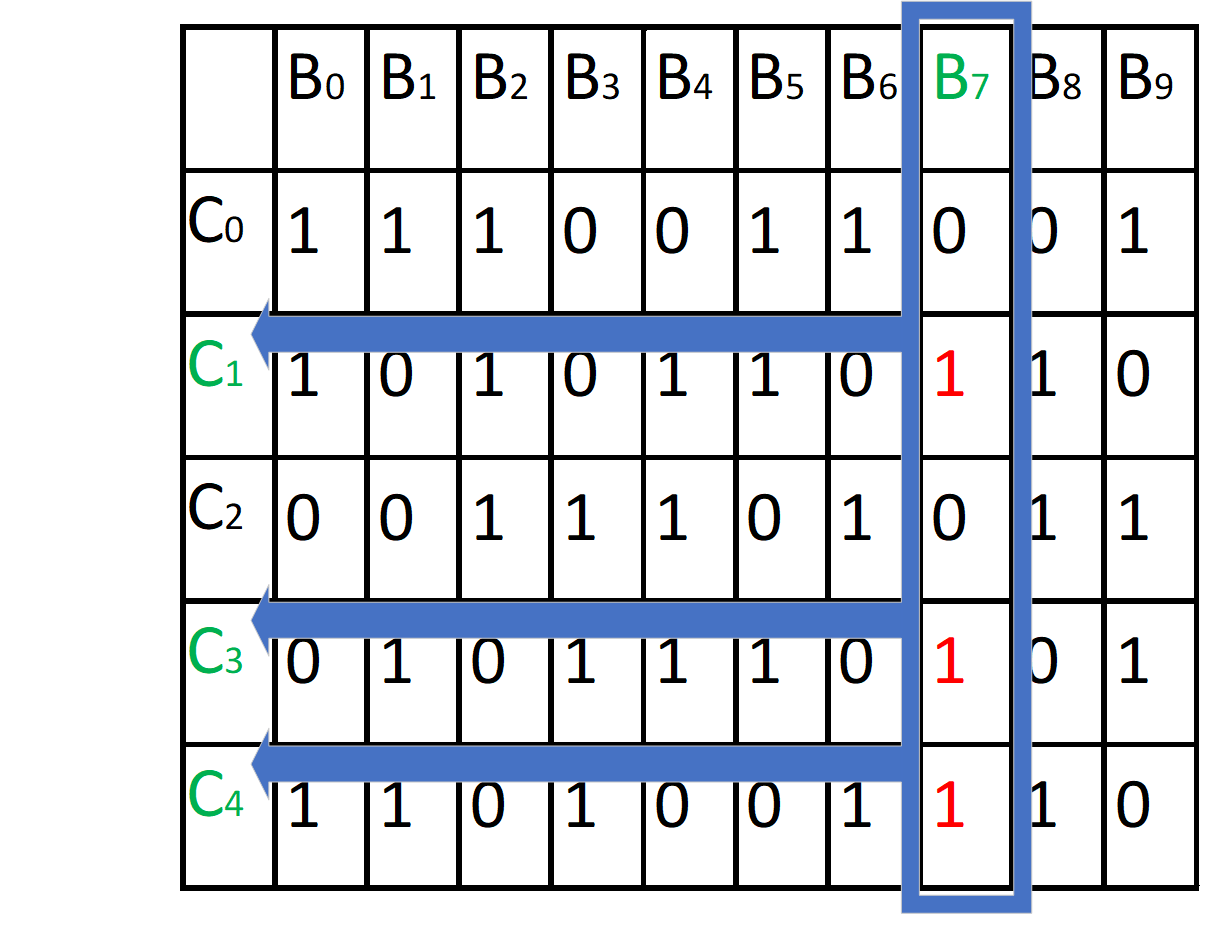}}
\caption{The sets $M_j$ are defined for every bit node and so correspond to the locations of non-zero elements in the \textit{columns} of the parity matrix.}
\label{fig:understandingMj} 
\end{figure}

We continue by examining the Min-Sum decoding algorithm as summarised here in Algorithm~\ref{alg:min-sum-graph-version}. 
A complete mathematical derivation, as well as other algorithms, can be found in \cite{Moon2005}.
In our context, it is more important to understand how the algorithm works and how it could be rephrased so it can be implemented in various ways.
As mentioned, an assignment of a binary sequence:
\begin{equation}
  c \in \{0,1\}^{||B||}
\end{equation}
is a codeword if and only if it satisfies all check nodes in $C$.
The Min-Sum algorithm, summarised in Algorithm~\ref{alg:min-sum-graph-version} and used in this work, uses this structure to pass messages between bit nodes and check nodes. Every bit node is assigned a real number representing Log-Likelihood-Ratio (LLR).
Its sign represents its logical value. In this work, strictly positive means logical "1", and non-positive means logical "0".
A bit node's magnitude represents the confidence of the bit node in its logical value, i.e., larger magnitude means more certainty.
We follow the notation from~\cite{Moon2005} with minor changes so that we can build on it in the following chapters.\
For a check node $c_i \in C$ denote by $N_i$ the subset of bit nodes that are connected to $c_i$, i.e.:
\begin{equation}
  N_i\; = \;\left\{b_j\, : \, H(i,j) = 1 \right\}
\end{equation}
the reader may note that these are determined by the locations of the non-zero entries in the $i'th$ row of the parity matrix $H$, 
and indeed this is demonstrated in Figure \ref{fig:understandingNi}.
Similarly, we denote by $M_j$ the set of check nodes that are connected to bit node $j$, i.e.:
\begin{equation}
  M_j\; = \;\left\{c_i\, : \, H(i,j) = 1 \right\}
\end{equation}
The reader may note this time that these correspond to the non-zero entries of the $j'th$ column in the parity matrix $H$.
Figure~\ref{fig:understandingMj} shows how to find the set $M_7$ using the parity matrix.
An observed vector $\boldsymbol{r}$ of real values serves as initial values for the bit nodes.
The task of the decoder is to find a value $\lambda_j$ for each bit node $b_j$ that maximises:
\begin{equation}
  P(\lambda_j\; |\; \boldsymbol{r},\, all\, checks\, involving\, b_j\, are\, satisfied)
\end{equation}
It should be noted that this is a computationally feasible approximation of the optimum decoder, i.e., a decoder that would seek a codeword $\boldsymbol{c}$ that maximises:
\begin{equation}
  P(\boldsymbol{c}\; | \;\boldsymbol{r},\, H\cdot \boldsymbol{c}\, =\, 0)
\end{equation}
The min-sum decoder starts by taking the received values $\boldsymbol{r}$ as a first estimation for the bit values, i.e.: 
\begin{equation}
  \lambda_j \; = \; \boldsymbol{r}(j) 
\end{equation}
The decoder continues by generating update values at the check nodes, followed by updating the values of the bit nodes.
Check node $c_i$ generates individual updates to each bit node $b_j$ which is a member of the set $N_i$, using the values $\lambda_j$ of the other bit nodes in the same set:
\begin{equation}
  \label{eq:eta_update}
  \eta_{i,j} = \min_{ b_k\in N_i,\, k \neq j} \left\{ |\lambda_k-\eta^{previous}_{i,k}|\right\} \cdot \prod_{\left\{k \neq j\; : b_k\in N_i\right\}}{sign(\lambda_k - \eta^{previous}_{i,k})}
\end{equation}
where $\eta^{previous}_{i,k}$ is the value calculated in the previous iteration by $c_i$ and sent to $b_k$. 
Where in the special case of the first iteration we define $\eta^{previous}_{k} = 0$.\\
The bit nodes update their values by summing all values calculated by the check nodes connected to them, as well as the initially observed information i.e.:
\begin{equation}
  \label{eq:lambda_j}
\lambda_j = \boldsymbol{r}(j) + \sum_{c_i \in M_j} \eta_{i,j}
\end{equation}
At this point, we can check if the algorithm converged to a binary code-word $b$ by "slicing" the inferred values vector $\lambda(j)$, i.e. setting:
\begin{equation}
  \label{eq:slice}
  \begin{split}
    b(i) &= 1\; if\; s(i) > 0\\
    b(i) &= 0\;  if\; s(i) \leq 0\\
  \end{split}
\end{equation}
\begin{algorithm}
	\SetAlgoLined
  \DontPrintSemicolon
  \KwIn{The parity matrix $H$, the received vector $\boldsymbol{r}$ and the maximum number of iterations $L$.}
	\KwOut{A binary vector $\boldsymbol{b}$, the iteration number at which the decoder stopped $k$, a status indication $isCodeword$.}
  Set $l = 0$\;
  Set $\forall{j} \,\lambda_j\,=\,\boldsymbol{r}(j)$\;
  set $\forall{i,j}\, \eta_{i,j}\,=\,0 $ \;
  Set isCodeword = False\;
	 \While{($k < L$) and ($\lnot $ isCodeword)}
	 {
    For every $c_j \in C$ calculate:\;
    $\eta_{i,j} = \min_{ b_k\in N_i,\, k \neq j} \left\{ |\lambda_k-\eta^{previous}_{i,k}|\right\} \cdot \prod_{\left\{k \neq j\; : b_k\in N_i\right\}}{sign(\lambda_k - \eta^{previous}_{i,k})} $\;
    For every $b_j \in C$ calculate:\;
    $\lambda_j = \boldsymbol{r}(j) + \sum_{c_i \in M_j} \eta_{i,j}$\;

    Calculate $\boldsymbol{b}$ according to equation~\ref{eq:slice}.\;
    
    \If{$H\cdot \boldsymbol{b} = \boldsymbol{0}$}
	 {
    set isCodeword = True\
   }
   $k = k + 1$\
   }
   \KwRet{isCodeword, $k$, $\boldsymbol{b}$}
	 \caption{Iterative Min-Sum decoding algorithm for binary LDPC codes \label{alg:min-sum-graph-version}}
	\end{algorithm}
\section{A map reduce architecture of the Min-Sum algorithm}
We begin this section with two observations regarding the Min-Sum algorithm~\ref{alg:min-sum-graph-version}. 
The first observation regards the term:
\begin{equation}
\min_{ b_k\in N_i,\, k \neq j} \left\{ |\lambda_k-\eta^{previous}_{i,k}|\right\}     
\end{equation}
in equation~\ref{eq:eta_update}:
\begin{observation}
  \label{ob:min}
The minimum taken in equation~\ref{eq:eta_update} does not need to be calculated for every $j$ in the set $N_i = \left\{b_j\; : \; H(i,j) = 1\right\}$.\\
Let: 
\begin{equation}
  min^i_0 = \min_{ b_k\in N_i} \left\{ |\lambda_k-\eta^{previous}_{i,k}|\right\}
\end{equation}

and:
\begin{equation}
  min_0^iLocation = arg\min_{\left\{ b_k\in N_i\right\}} \left\{ |\lambda_k-\eta^{previous}_{i,k}|\right\}
\end{equation}
and finally:\\
\begin{equation}
  min^i_1 = \min_{\left\{ b_k\in N_i,\; k\neq min0Location\right\}} \left\{ |\lambda_k-\eta^{previous}_{i,k}|\right\}
\end{equation}
then there are two cases: \\
\textbf{Case 1:} $j \neq min^i_0Location$ and hence \\
\begin{equation}
\min_{\left\{ b_k\in N_i,\, k \neq j\right\}} \left\{ |\lambda_k-\eta^{previous}_{i,k}|\right\}\; =\; min0
\end{equation}

\textbf{Case 2:} $j = min0Location$ and hence\\
\begin{equation}
\min_{\left\{ b_k\in N_i,\, k \neq j\right\}} \left\{ |\lambda_k-\eta^{previous}_{i,k}|\right\}\; = \;min1
\end{equation}
\end{observation}
Observation~\ref{ob:min} can be found in \cite{Yuan2017}.

The second observation is similar in nature to observation~\ref{ob:min}, 
in that, we seek to reduce the number of numerical operations.
The second term of equation~\ref{eq:eta_update}, i.e.: 
\begin{equation}
\prod_{\left\{k \neq j\; : b_k\in N_i\right\}}{sign(\lambda_k - \eta^{previous}_{i,k})}
\end{equation}
requires a product of signs.
\begin{observation}
  \label{ob:sign}
   Assuming the function $sign(\cdot)$ returns either $1$ or $-1$ (but not $0$), we have:
   \begin{equation}
    \begin{split}
      \prod_{\left\{k \neq j\; : b_k\in N_i\right\}}{sign(\lambda_k - \eta^{previous}_{i,k})} = \\
      = \prod_{\left\{b_k\in N_i\right\}}{sign(\lambda_k - \eta^{previous}_{i,k})} \times \frac{1}{sign(\lambda_j - \eta^{previous}_{i,j})}\\  
      = \prod_{\left\{b_k\in N_i\right\}}{sign(\lambda_k - \eta^{previous}_{i,k})} \times sign(\lambda_j - \eta^{previous}_{i,j})
    \end{split}
   \end{equation}
   where the last equality stems from the fact that $\frac{1}{1} = 1$ and $\frac{1}{-1} = -1$.
\end{observation}

The importance of observation~\ref{ob:min} and observation~\ref{ob:sign}, other than the apparent reduction of mathematical operations, 
is that they allow us to reformulate the Min-Sum algorithm using reduction operation on matrices.
Reduction operations lend themselves to Single Instruction Multiple Threads (SIMT) architectures, which are common to Graphics Processing Unit (GPU). Examples of reductions are the functions $min(\cdot), sum(\cdot), product(\cdot)$, and implementation of reduction operations on GPUs are explained in \cite{Kirk2016} and \cite{Harris2007}.

In this section we will restate the components $\eta$ and $\lambda$ of the Min-Sum algorithm~\ref{alg:min-sum-graph-version} 
so that a single iteration could be easily seen as a sequence of fan-out (map) and reduction operations.

If we define $\eta$ as a matrix, i.e.:
\begin{equation}
    \eta(i,j) \triangleq \eta_{i,j} 
\end{equation}
Then equation~\ref{eq:lambda_j} takes the form:
\begin{equation}
  \begin{split}
\lambda_j & = \boldsymbol{r}(j) + \sum_{c_i \in M_j} \eta_{i,j}\\
          & = \boldsymbol{r}(j) + \sum_{c_i \in M_j} \eta(i,j)
  \end{split}
\end{equation}
If we were to make sure that in the above equation we have $\eta(i,j) = 0$ whenever $c_i \notin M_j$, then we would obtain:
\begin{equation}
  \begin{split}
  \lambda_j & = \boldsymbol{r}(j) + \sum_{c_i \in M_j} \eta(i,j) \\
            & = \boldsymbol{r}(j) + \sum_{c_i \in M_j} \eta(i,j) + 0 \\
            & = \boldsymbol{r}(j) + \sum_{c_i \in M_j} \eta(i,j) + \sum_{c_i \notin M_j} \eta(i,j) \\
            & = \boldsymbol{r}(j) +\sum_{i} \eta(i,j)
  \end{split}
\end{equation}
We make sure of that by setting $\eta(i,j) = 0$ whenever $H(i,j) = 0$, i.e.:
\begin{equation}
  H(i,j) = 0 \implies \eta(i,j) = 0
\end{equation}
This allows for the implementation of the summation part of equation~\ref{eq:lambda_j} as a reduction operation, rather than needing control logic.
Next we turn our attention to equation~\ref{eq:eta_update}. 
Plugging in observations \ref{ob:min} and \ref{ob:sign} as well as our definition of $\eta$ as a matrix we obtain:

\begin{multline}
\eta_{i,j} =\\
\min_{ b_k\in N_i,\, k \neq j} \left\{ |\lambda_k-\eta^{previous}_{i,k}|\right\} \cdot  \prod_{\left\{k \neq j\; : b_k\in N_i\right\}}{sign(\lambda_k - \eta^{previous}_{i,k})}=\\
\min_{ b_k\in N_i,\, k \neq j} \left\{ |\lambda_k-\eta^{previous}_{i,k}|\right\}\cdot \\
\cdot\prod_{k\; : \left\{b_k\in N_i\right\}} {sign(\lambda_k - \eta^{previous}_{i,k})} 
\cdot sign(\lambda_j - \eta^{previous}_{i,j}) =\\
(\delta_{j, min^i_0Location} \cdot min^i_1) + ((1-\delta_{j, min^i_0Location})\cdot min^i_0) \cdot\\
\cdot \; \prod_{\left\{b_k\in N_i\right\}}{sign(\lambda_k - \eta^{previous}_{i,k})} \cdot sign(\lambda_j - \eta^{previous}_{i,j})
\end{multline}

We have already set $\eta(i,j) = 0$ whenever $H(i,j) = 1$. Assume that:
\begin{displaymath}
  sign(0) = 1
\end{displaymath}

then we have the following:
\begin{align}
\prod_{\left\{b_k\in N_i\right\}}{sign(\lambda_k - \eta^{previous}_{i,k})} \cdot sign(\lambda_j - \eta^{previous}_{i,j}) =\\
     1\; \cdot\; \prod_{\left\{b_k\in N_i\right\}}{sign(\lambda_k - \eta^{previous}_{i,k})} \cdot sign(\lambda_j - \eta^{previous}_{i,j}) \\
    = \prod_{\left\{b_k\notin N_i\right\}}{sign(\lambda_k - \eta^{previous}_{i,k})} \\
    \cdot \prod_{\left\{b_k\in N_i\right\}}{sign(\lambda_k - \eta^{previous}_{i,k})}  \\
    \cdot sign(\lambda_j - \eta^{previous}_{i,j})\\
    = \prod_{i}{sign(\lambda_k - \eta^{previous}_{i,k})} \cdot sign(\lambda_j - \eta^{previous}_{i,j})
\end{align}

and the term 
\begin{equation}
\prod_{i}{sign(\lambda_k - \eta^{previous}_{i,k})}
\end{equation} is a reduction operation.
\noindent We are now ready to restate the Min-Sum algorithm in map and reduce form.\\
\noindent We begin with the inputs:
\begin{enumerate}
  \item The parity matrix $H$ with dimensions $m\times n$
  \item A set of values $\boldsymbol{r}$ is observed at the receiver
  \item The maximal number of iterations is $L$
\end{enumerate}
The first iteration of the Min-Sum algorithm starts as follows:
We first map the vector of observation $\boldsymbol{r}$ to a matrix by taking as many copies as there are check nodes, $m$, followed by masking with $H$ and subtracting $\eta$:
\begin{equation}
  \label{eq:firstIterationFanOut}
  \lambda (i,j) \triangleq \boldsymbol{r}(j) \cdot H(i,j)
\end{equation}
We proceed by a reduction of $\lambda$ to four vectors of size $m$; $min_0$, $min_1$ and $min_0Location$ defined as follows:
\begin{equation}
    min_0(i) = \min_{0 \leq j < m} \left\{ |lambda(i,j)| : H(i,j) = 1 \right\}
\end{equation}
and
\begin{equation}
    min_0Location(i) = arg\min_{0 \leq j < m} \{ |lambda(i,j)| : H(i,j) = 1 \}
\end{equation}
as well as:
\begin{equation}
        min_1(i) = \min_{0 \leq j < m, \; i\neq min_0Location(i)} \left\{ |\lambda(i,j)| : H(i,j) = 1 \right\} 
\end{equation}
and finally:
\begin{equation}
    sgn(i) = \displaystyle\prod_{ \left\{ 0 \leq j < m \right\} } sign(\lambda(i,j))\\
\end{equation}
It should be noted that the first three vectors could be calculated at once using a sort function truncated to the two smallest elements.
Furthermore, the sign function here satisfies $sign(0) = 1$.
We now have all the information we need to update the matrix $\eta$ as follows:
\begin{equation}
  \begin{split}
    \label{eq:etaCalculation}
    \eta(i,j) =& H(i,j) \cdot \\
               & sign(\lambda(i,j)) \cdot sgn(i) \cdot\\
               & (\delta_{j,min_0Location(i)}\cdot min_0(i) + \\
               & (1-\delta_{j,min_0Location(i)})\cdot min_1(i))
  \end{split}
\end{equation}
The next part is the "sum" part of the Min-Sum algorithm, 
which we perform as a column-wise sum reduction of the matrix $\eta$, and add to it the initial observations vector $\boldsymbol{r}$:
\begin{equation}
  \begin{split}
    \label{eq:sCalculation}
    s(j) = \displaystyle\sum_{ \left\{ 0 \leq j < n \right\} } \eta(i,j) + r(j)\\
  \end{split}
\end{equation}
At this point, we check if the algorithm converged to a binary code-word $\boldsymbol{b}$ by slicing the vector $\boldsymbol{s}$ as in equation~\ref{eq:slice}.
i.e. setting:\begin{equation}
  \begin{split}
    b(i) &= 1\; if\; s(i) > 0\\
    b(i) &= 0\;  if\; s(i) \leq 0
  \end{split}
\end{equation}

And then checking whether $\boldsymbol{b}$ is in the kernel of $H$, i.e.:
\begin{equation}
  \begin{split}
    H\cdot \boldsymbol{b} \;\;\; ?= \;\;\; \boldsymbol{0}
  \end{split}
\end{equation}
where the last equation is over the binary field, and $\boldsymbol{0}$ is the all $0s$ vector.\\
If $\boldsymbol{b}$ is indeed in the kernel of $H$, then $\boldsymbol{b}$ is a code-word and we can stop the algorithm.
Otherwise, we increase the iteration counter $k$ by $1$:
\begin{equation}
    k := k + 1
\end{equation}
The algorithm proceeds iteratively, with a single difference in subsequent iterations. 
We replace equation~\ref{eq:firstIterationFanOut} with the following update of the matrix $\lambda$, 
based on the vector $\boldsymbol{s}$ instead of $\boldsymbol{r}$, and reducing the formerly calculated $\eta$:
\begin{equation}
  \lambda (i,j) \triangleq \boldsymbol{s}(j) \cdot H(i,j) - \eta(i,j)
\end{equation}
We note that this could be incorporated into the first iteration (setting $\eta$ initially to an $m\times n\;\; 0s$ matrix as well as $\boldsymbol{s}$ to be $\boldsymbol{r}$).\\

\begin{algorithm}
	\SetAlgoLined
  \DontPrintSemicolon
  \label{alg:min-sum}
  \KwIn{The parity matrix $H$, the received vector $\boldsymbol{r}$ and the maximum number of iterations $L$.}
	\KwOut{A binary vector $\boldsymbol{b}$, the iteration number at which the decoder stopped $k$, a status indication $isCodeword$.}
  Set $\eta(i,j) = 0$ for all $i,j$.\;
  Set $\lambda(i,j) = r(i) \cdot H(i,j)=1$.\; 
  Set isCodeword = False\;
	 \While{($k < L$) and ($\lnot $ isCodeword)}
	 {
    calculate the matrix $\eta$ according to equation~\ref{eq:etaCalculation}.\;
	  calculate the vector $\boldsymbol{s}$ according to equation~\ref{eq:sCalculation}.\;
    calculate $\boldsymbol{b}$ according to equation~\ref{eq:slice}.\;
    
    \If{$H\cdot \boldsymbol{b} == \boldsymbol{0}$}
	 {
    set isCodeword = True\
   }
   $k := k + 1$\
   }
   \KwRet{isCodeword, $k$, $\boldsymbol{b}$}
	 \caption{Iterative Min-Sum decoding algorithm for binary \LDPC{codes}}
	\end{algorithm}

\section{CUDA kernels}
In this section, we explain how to break the algorithm further so it could be implemented using CUDA kernels and device functions using NUMBA \cite{Lam2015}. By convention, variables that end with $\_device$ exist on the GPU memory, whereas $\_host$ signifies that they reside on the host CPU. The given here \ref{lst:eval} shows the signature of every kernel used. Parameters such as Blocks Per Grid (BPG) and Threads Per Block (TPB) have to be calculated before hand, and experience shows that fine-tuning them separately for each kernel may improve performance. The full version of this code can be found in \cite{cudaLDPC}.

\begin{lstlisting}[caption={Architecture of the Min-Sum decoder as a sequence of CUDA kernels}, label={lst:eval}, language=Python]
# Move assets from host to CUDA device
fromChannel_device = cuda.to_device(fromChannel_host)    
softVector_device = cuda.to_device(softVector_host)
iterator = 0
# Set variable to False
isCodeword = False
# Repeat the vector softVector_device which size is [1, numberOfBitNodes] to create a matrix which dimension is [numberOfCheckNodes, numberOfBitNodes]
maskedFanOut[BPG_DIM1, TPB](parityMatrix_device, softVector_device, matrix_device)
#Slice (make a binary decision) out of the values of softVector_device   
slicerCuda[BPG_DIM1, TPB](softVector_device, binaryVector_device)
# Set isCodewordVector_device to zero
resetVector[DIM1, 1](isCodewordVector_device)
# Calculate the product of the parity matrix and the sliced (binary) values
calcBinaryProduct2[BPG_BINARY_CALC, TPB_BINARY_CALC](parityMatrix_device, binaryVector_device, isCodewordVector_device)
# Apply module 2 to the result
mod2Vector[BPG, TPB](isCodewordVector_device)
#Check if all entries are 0, and store the result in location 0:
checkIsCodeword[BPG_DIM0, TPB](isCodewordVector_device, result_device)
if result_device[0] == 0 :
    isCodeword = True
while (iterator < numberOfIterations and not isCodeword):
    # Find (in every row) the smallest element, its location, and the second smallest element. We also use this kernel to find the number of negative values.
    findMinimaAndNumberOfNegatives[BPG_DIM0, TPB](parityMatrix_device, matrix_device, smallest_device, secondSmallest_device, locationOfMinimum_device)
    # Next we calculate the (extended) sign from the number of negatives, -1 for odd and 1 for even
    numberOfNegativesToProductOfSigns[BPG_DIM0, TPB](numberOfNegatives_device,productOfSigns_device)  
    # Find the two smallest elements (in rows, concurrently) and their location
    locateTwoSmallestHorizontal2DV2[BPG_LOCATE_TWO_SMALLEST_HORIZONTAL_2D, TPB_LOCATE_TWO_SMALLEST_HORIZONTAL_2D](matrix_device, parityMatrix_device, smallest_device, secondSmallest_device, locationOfMinimum_device)
    # Calculate the sign of each row
    signReduceHorizontal[DIM0, 1](matrix_device, productOfSigns_device)  
                    produceNewMatrix2D[BPG_PRODUCE_NEW_MATRIX_2D, TPB_PRODUCE_NEW_MATRIX_2D](parityMatrix_device, matrix_device, smallest_device, secondSmallest_device, locationOfMinimum_device, productOfSigns_device, newMatrix_device)
    # Sum vertically. Notice that the result of matrix summation using the cuda kernel does not have to be exactly as the numpy sum.
    matrixSumVertical[BPG_DIM1, TPB](newMatrix_device, softVector_device)
    cudaPlusDim1[BPG_DIM1, TPB](softVector_device,fromChannel_device)
    
    # Check if the result is a codeword, similar to what we did before entering decoding.
    slicerCuda[BPG_DIM1, TPB](softVector_device, binaryVector_device)
    resetVector[1022, 1](isCodewordVector_device)
    calcBinaryProduct2[BPG_BINARY_CALC, TPB_BINARY_CALC](parityMatrix_device, binaryVector_device, isCodewordVector_device)
    mod2Vector[BPG_16, TPB](isCodewordVector_device)
    checkIsCodeword[BPG_DIM0, TPB](isCodewordVector_device, result_device)
    if result_device[0] == 0:
        isCodeword = True
    # Fan out the last LLR values to be used again
    maskedFanOut[BPG_DIM1, TPB](parityMatrix_device, softVector_device, matrix_device)
    # Reduce:  matrix_device - newMatrix_device
    cudaMatrixMinus2D[BPG_MATRIX_MINUS_2D, TPB_MATRIX_MINUS_2D](matrix_device, newMatrix_device) 
    iterator = iterator + 1
slicerCuda[BPG_DIM1, TPB](softVector_device, binaryVector_device)
result_device[1] = 0
numberOfNonZeros[BPG_DIM1, TPB](binaryVector_device, result_device)
result_host = result_device.copy_to_host()
berDecoded = result_host[1]
\end{lstlisting}

\section{Related work}
The implementation of a message-passing decoder developed in this work uses the min-sum decoding algorithm described in (for example) \cite{Moon2005} and \cite{Li2016a}.
A decoder architecture for 60GHz WiFi (802.11ad) and Bluetooth (802.153c) can be found in \cite{Weiner2011}.
The work there provides micro architecture hardware schematics as well as an estimate of power consumption.
Some insights on Graphical Processing Unit (GPU) implementation could be found in \cite{Yuan2017},
where a 4.7 Gbps decoder is implemented on a GeForce 1080Ti.
The \LDPC used in \cite{Yuan2017} was the same as the one used in WiFi standard 802.11n.
A similar work but specific to the codes used in the Digital Video Broadcasting (DVB-S2) \cite{DVB2003} is presented in \cite{Kun2018}.
For the codes used in 5G, a comparison between an FPGA and a GPU implementation is given in \cite{Aronov2019}, 
and a 150Mbps software decoder is reported in \cite{Kaltenberger2019}.
In this work we use CUDA kernels to implement reduction functions. The reader would find such reduction functions explained in \cite{Harris2007}, and \cite{Kirk2016}.
\section{Experimental resulsts}
\begin{figure}[h]
  \centering
  \includegraphics[width = 0.9\linewidth]{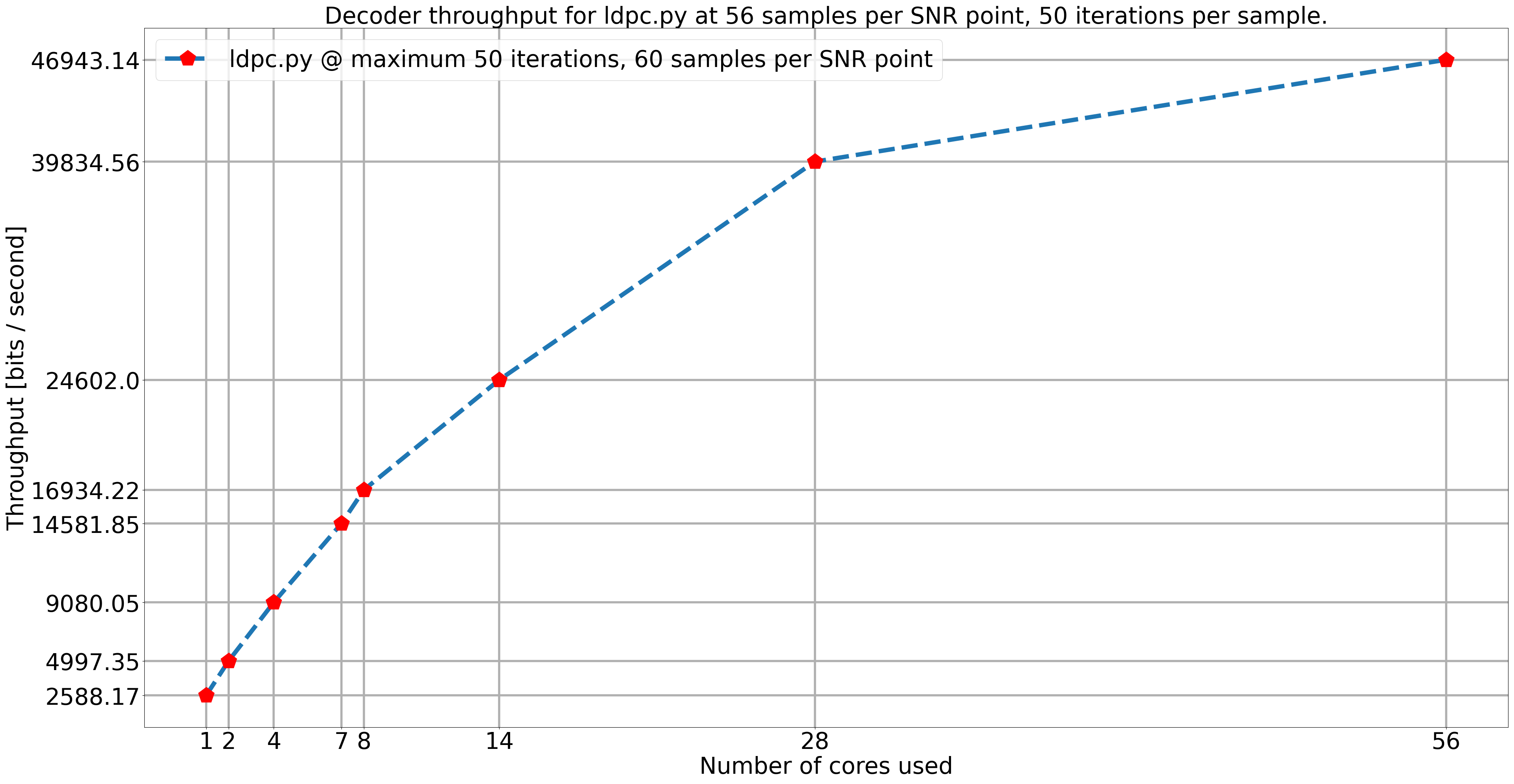}
  \caption{Scaling results of our implementation of a 
  Min-Sum decoder that relies on the Tanner graph representation of 
  an \LDPC{code}. For each fixed number of cores, $56$ samples of each of the SNR points: [3.0 ,3.2, 3.4] 
  were chosen at random and up to $50$ Min-Sum iterations were allowed.\label{fig:ldpcPyScaling}}
\end{figure}

\begin{figure}[h]
  \centering
  \includegraphics[width = 0.9\linewidth]{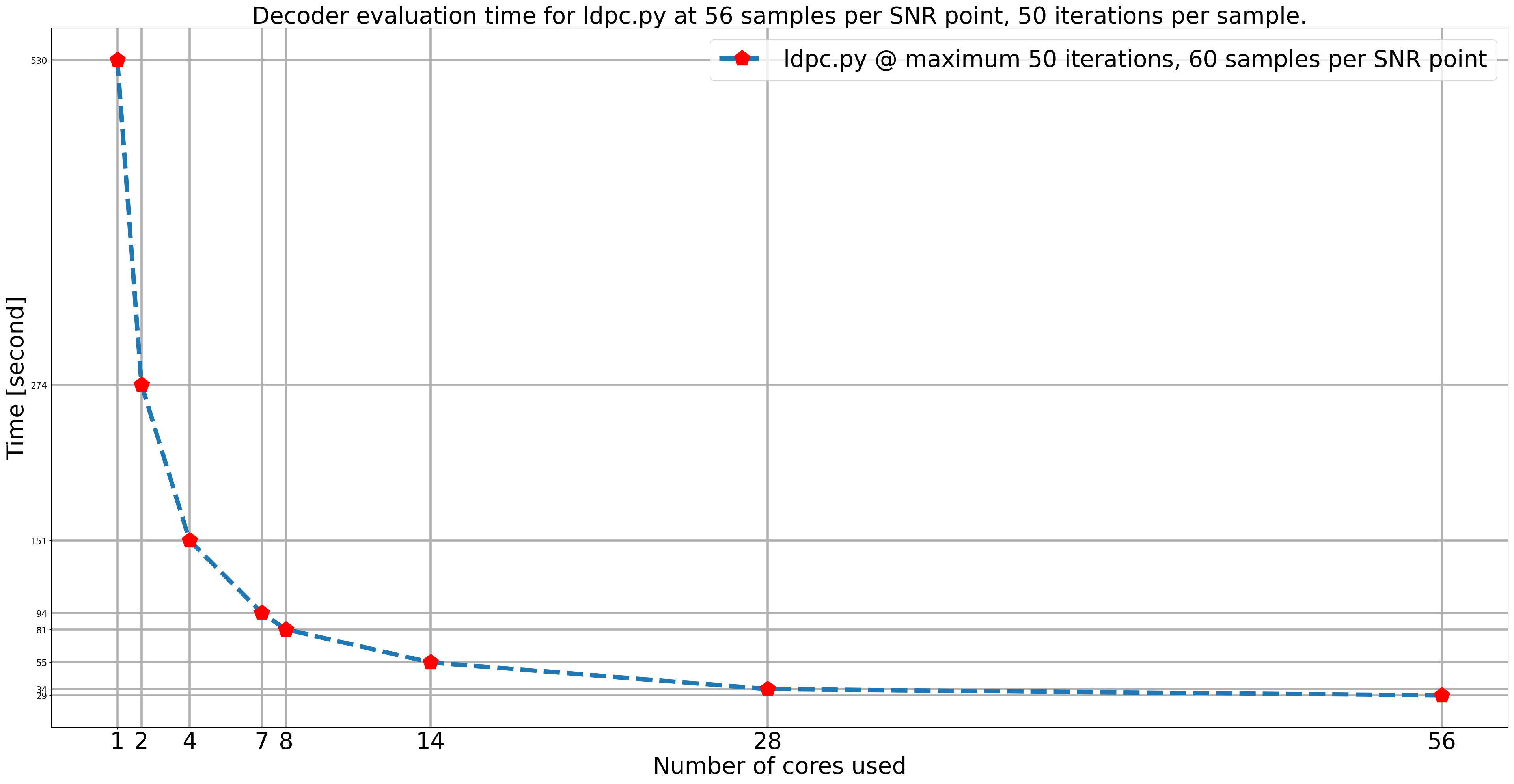}
  \caption{Scaling results of our implementation of a 
  Min-Sum decoder showing the total amount of time is takes to: sample $56$ times each of the SNR points: [3.0 ,3.2, 3.4] 
  and decode using up to $50$ Min-Sum iterations.\label{fig:ldpcScalingTime}}
\end{figure}

We use a Quaci-Cyclic 8176,1022 code from \cite{CCSDS2007} to test decoder throughput. 
For comparison, we implemented a pure CPU-oriented of the Min-Sum decoder in Python with as much usage of the Numba package \cite{Lam2015} to compile code directly to the machine language as possible.
This is a good approach if the content of the sparse parity matrix $H$ is known, and many of the calculations could be discarded beforehand. Figure~\ref{fig:ldpcPyScaling}, 
however, shows its scaling limitations for decoding the benchmark parity matrix taken from \cite{CCSDS2007} as a function of the number of cores utilised.
The reader may observe that although the throughput increases, there are 
diminishing returns. Figure~\ref{fig:ldpcScalingTime} shows the same evaluation, this time with the y-axis signifying time in seconds. 

Next, we turn to examine our formulation of the Min-Sum algorithm implemented using CUDA kernels, on a node equipped with four NVIDIA A100 GPUs per node.

\begin{figure}[h]
  \centering
  \includegraphics[width = 0.9\linewidth]{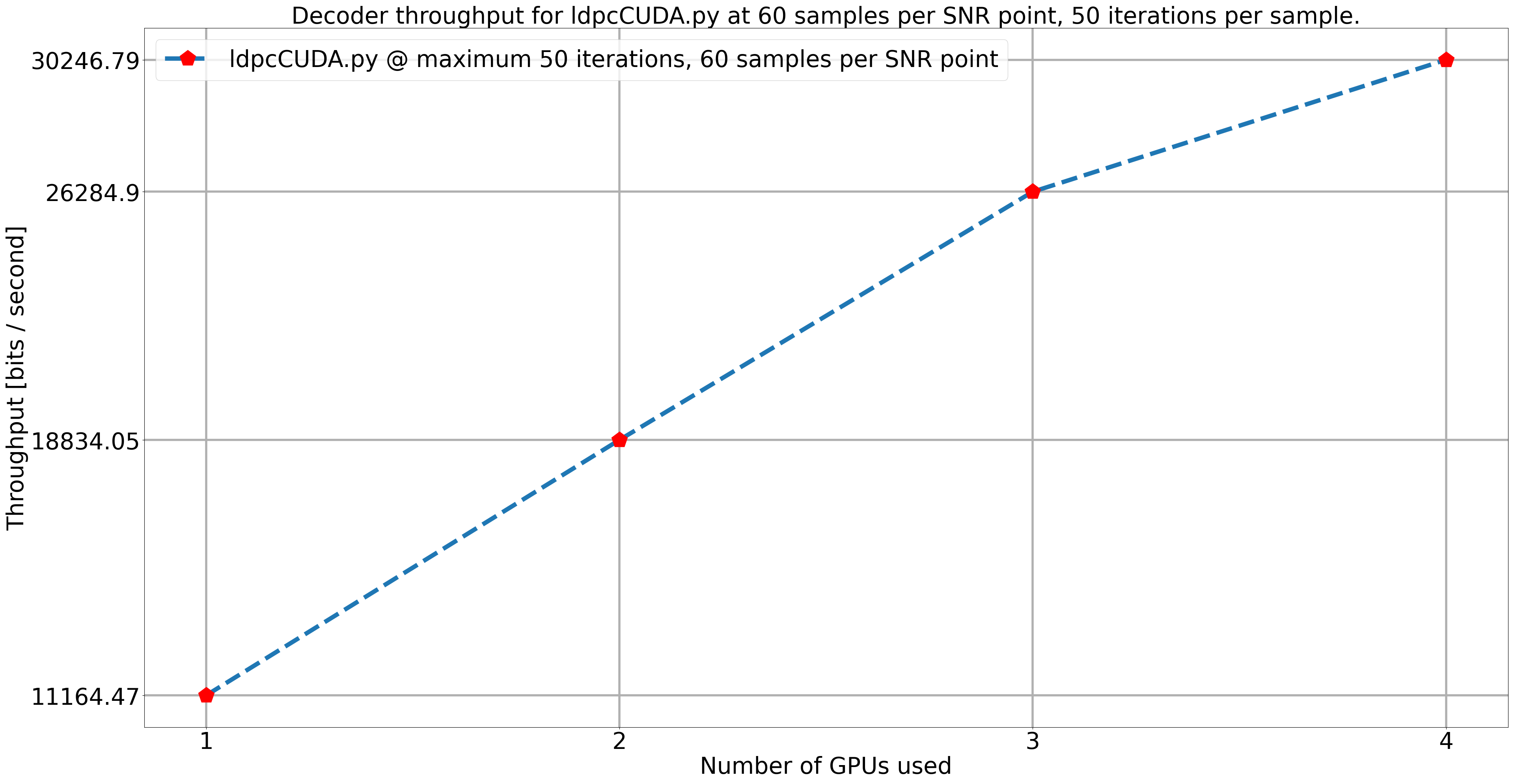}
  \caption{GPU formulated Min-Sum decoder. For each fixed number of cores, $60$ samples of each of the SNR points: [3.0 ,3.2, 3.4] 
  were chosen at random and up to $50$ Min-Sum iterations were allowed. Termination was checked for every $6$ iterations.\label{fig:ldpcCUDAScalingThroughput}}
\end{figure}
\begin{figure}[ht]
  \centering
  \includegraphics[width=0.9\linewidth]{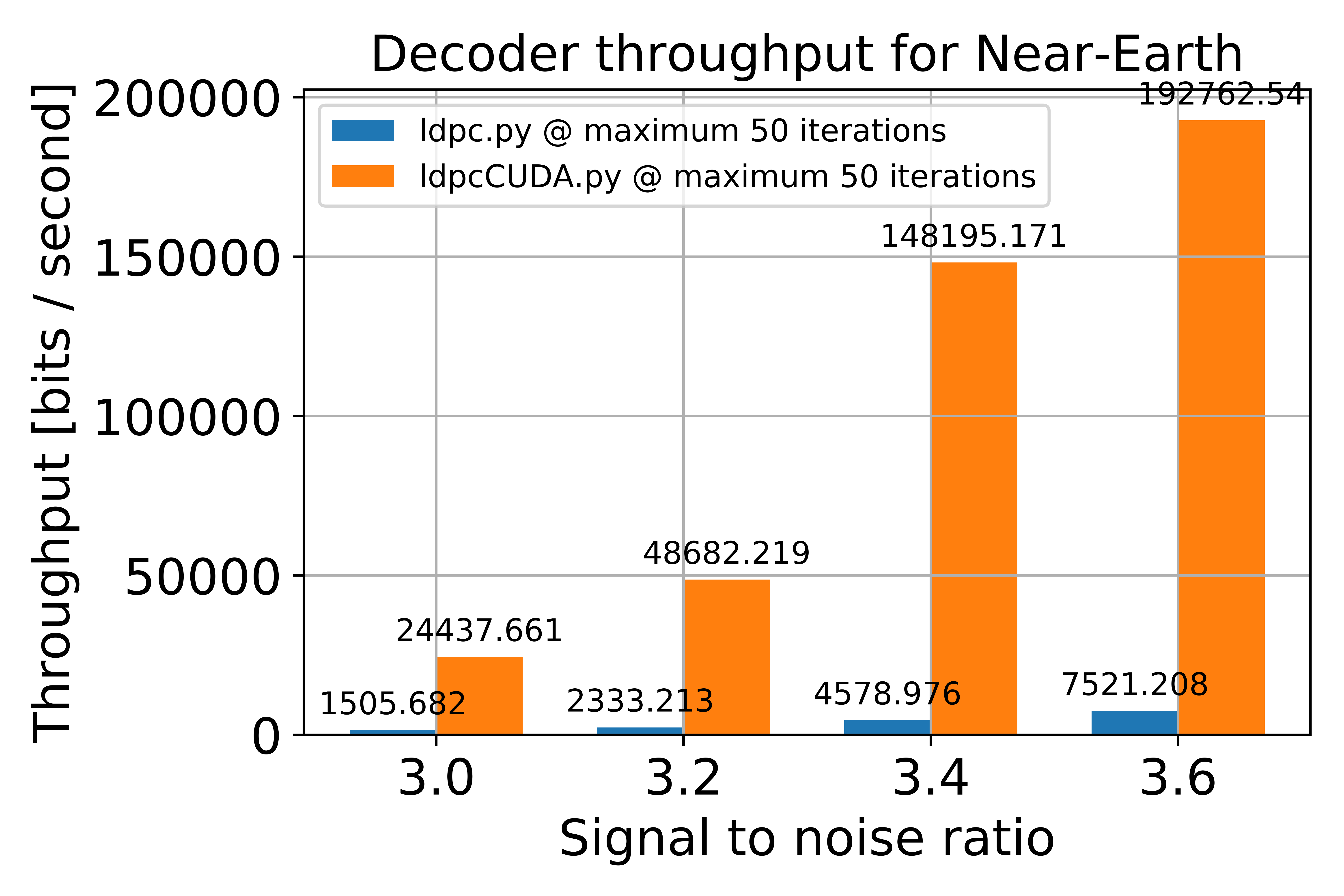}
  \caption{Throughput as a function of SNR when the parity matrix is taken from \cite{CCSDS2007}. 
  At each SNR point, fifty samples of AWGN noise were generated and added to the all-zero codeword. The indicated throughput was calculated as 
  $\frac{samples \times 8176\, bits}{Processing\, time\, in\, seconds}$.
  \label{fig:min-sum-cuda-throughput}}
\end{figure}
\begin{figure}
  \centering
  \includegraphics[width = 0.8\linewidth]{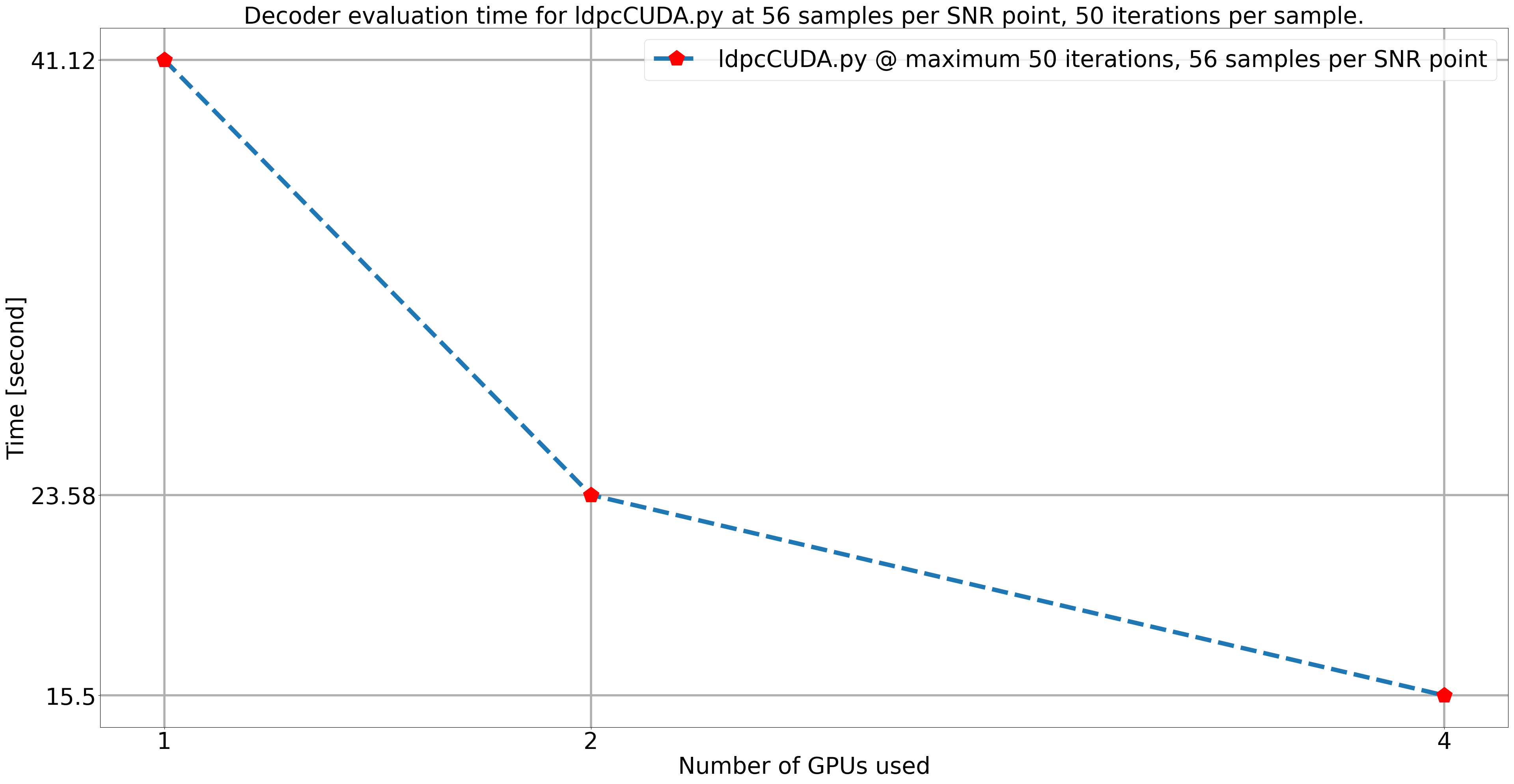}
  \caption{Scaling results of our implementation of a 
  GPU formulated Min-Sum decoder showing the total amount of time is takes to: sample $60$ times each of the SNR points: [3.0 ,3.2, 3.4] 
  and decode using up to $50$ Min-Sum iterations.\label{fig:ldpcCUDAScalingTime}}
\end{figure}

\noindent The reader may note we used $60$ samples per SNR point in Figure~\ref{fig:ldpcCUDAScalingThroughput} as opposed to $56$ in Figure~\ref{fig:ldpcPyScaling}.
This was in order to capture the performance on $1,2,3$ and $4$ GPUs. 
For evaluation time to be compared between Figure~\ref{fig:ldpcScalingTime}
and Figure~\ref{fig:ldpcCUDAScalingTime} we reduced the number of samples back to $56$ to match both figures, neglecting the 3-GPU data point.
This brings us to the throughput results in Figure~\ref{fig:min-sum-cuda-throughput}. 
Independent of how efficient our CPU-oriented implementation may be, we provide a profile of our CUDA-kernels-based code in table \ref{tab:cudaKernelsProf}.

\begin{table}
	\centering
  \begin{tabular}{ |p{1cm}|p{2cm}|p{4.5cm}|  }
  \hline
Time($\%$)&  Total Time (ns)&  CUDA kernel name\\
\hline
   $68.6$&$   30,782,512,889 $&   findMinimaAndNumberOfNegatives\\
   $13.3$&$    5,970,164,807 $&   signReduceHorizontal\\
   $ 7.4$&$    3,315,661,996$&    matrixSumVertical\\
   $ 6.3$&$    2,812,575,858$&    maskedFanOut\\
   $ 1.5$&$      667,298,761$&    locateTwoSmallestHorizontal2DV2\\
   $ 1.3$&$      571,863,673$&    produceNewMatrix2D\\
   $ 0.9$&$      406,282,038$&    cudaMatrixMinus2D\\
   $ 0.4$&$      173,104,002$&    calcBinaryProduct2\\
   $ 0.1$&$       25,418,451$&    cudaPlusDim1\\
   $ 0.1$&$       24,269,465$&    resetVector\\
   $ 0.1$&$       24,177,579$&    numberOfNegativesToProductOfSigns\\
   $ 0.1$&$       23,348,399$&    slicerCuda\\
   $ 0.0$&$       21,828,707$&    checkIsCodeword\\
   $ 0.0$&$       21,406,907$&    mod2Vector\\
   $ 0.0$&$          840,889$&    numberOfNonZeros\\

  \hline

  \end{tabular}
  \caption{Table of time spent in each CUDA kernel used. The total run time was $48.31$ seconds with $60$ iterations of the Min-Sum decoder, and $50$ samples at each of the following SNR points: $3.0,3.2,3.4,3.6$. A valid codeword was checked for every $6$ iterations.].\label{tab:cudaKernelsProf}}
  \end{table}
\noindent While profiling the code running on the GPU, an interesting observation emerged: over $95\%$ of the time spent in API functions was used to copy memory from the GPU and to the host.
At the same time, these consume several orders of magnitude smaller memory than host to device memory copy as seen in Table~\ref{tab:memoryFootprint}.
  \begin{table}
    \centering
    \begin{tabular}{ |p{1cm}|p{2.5cm}|p{3cm}|  }
      \hline

      Time($\%$)&  Total Time (ns)&  API call name\\
      \hline
        $96.5$&   $31,300,527,019$&   cuMemcpyDtoHv2\\
          $2.1$&      $673,476,248$&   cuLaunchKernel\\
          $1.4$&      $445,250,879$&   cuMemcpyHtoDv2\\
          $ < 0.1$&        $9,827,377$&   cuMemAllocv2\\
          $ < 0.1$&        $5,016,936$&   cuMemFreev2\\
          $ < 0.1$&        $2,164,764$&   cuModuleLoadDataEx\\
          $ < 0.1$&        $1,738,263$&   cuLinkComplete\\
          $ < 0.1$&          $790,449$&   cuModuleUnload\\
          $ < 0.1$&          $736,754$&   cuLinkCreatev2\\
          $ < 0.1$&           $32,852$&   cuMemGetInfov2\\
          $ < 0.1$&           $22,101$&   cuLinkDestroy\\
          $ < 0.1$&           $ 3,206$&  cuInit\\
          \hline
  \end{tabular}
  \caption{API calls that were used in order to run the code on a GPU. The total run time was $48.31$ seconds with $60$ iterations of the Min-Sum decoder, and $50$ samples at each of the following SNR points: $3.0,3.2,3.4,3.6$. A valid codeword was checked for every $6$ iterations.\label{tab:cudaAPIProf}}
  \end{table}
  This was investigated by changing the frequency at which the decoder checks for convergence (isCodeword == 0), and proved to have an affect of the number of calls.
  More generally, the control logic around the CUDA kernels required communication with the host. Changing the frequency, however, did not contribute to the performance of the code when running with cuda toolkit 11.2.
  \begin{table}[ht]
    \centering
    \begin{tabular}{ |p{2cm}|p{1.5cm}|p{3cm}|  }
      \hline
  Total&     Operations&          Operation\\
  \hline
       74.234&       4,724 &  CUDA memcpy DtoH\\
  105,642.695&         374 &  CUDA memcpy HtoD\\
  \hline
\end{tabular}
\caption{CUDA Memory Operation Statistics (by size in KiB) \label{tab:memoryFootprint}}
\end{table}

\bibliographystyle{plainurl}
\bibliography{RL-CODES.bib}
\end{document}